\documentclass[%
 aip,cha,
 floatfix,
 preprintnumbers,
 reprint,
 groupedaddress,
 amsmath,amssymb,
]{revtex4-1}

\newcommand*{\papertitle}{Non-linear spin Seebeck effect due to spin-charge interaction in graphene}

\usepackage{graphicx}% Include figure files
\usepackage{bm}% bold math
\usepackage[breaklinks=true]{hyperref}
\hypersetup{bookmarksnumbered, pdfpagemode=UseOutlines, 
pdfauthor={I.\ J.\ Vera-Marun}, 
pdftitle={\papertitle}, 
pdfdisplaydoctitle, 
colorlinks=true, citecolor=blue, filecolor=blue, linkcolor=blue, urlcolor=blue}

\begin{document}

\title{\papertitle}
%\thanks{A footnote to the article title}%

\author{I.\ J.\ \surname{Vera-Marun}}
\email[e-mail: ]{I.J.Vera.Marun@rug.nl} %\homepage[]{} \thanks{}
\author{V.\ Ranjan}
\author{B.\ J.\ \surname{van Wees}}
\affiliation{Physics of Nanodevices, Zernike Institute for Advanced Materials, University of Groningen, The Netherlands}

%\date{\today}

\begin{abstract}
\end{abstract}

%\pacs{72.25.Hg, 72.80.Vp, 75.76.+j, 85.75.-d}
%\keywords{spintronics, graphene}
\preprint{Preprint v5 (with supplement). Dated: \today}

\maketitle

\noindent \textbf{%
The abilities to inject and detect spin carriers are fundamental for research on transport and manipulation of spin information \cite{uti_spintronics:_2004, chappert_emergence_2007}. Pure electronic spin currents have been recently studied in nanoscale electronic devices using a non-local lateral geometry, both in metallic systems \cite{jedema_electrical_2001} and in semiconductors \cite{lou_electrical_2007}. To unlock the full potential of spintronics we must understand the interactions of spin with other degrees of freedom, going beyond the prototypical electrical spin injection and detection using magnetic contacts. Such interactions have been explored recently, for example, by using spin Hall \cite{valenzuela_direct_2006, uchida_observation_2008, abanin_giant_2011-3} or spin thermoelectric effects \cite{uchida_observation_2008, slachter_thermally_2010-1, le_breton_thermal_2011}. Here we present the detection of non-local spin signals using non-magnetic detectors, via an as yet unexplored non-linear interaction between spin and charge. In analogy to the Seebeck effect \cite{cutler_observation_1969}, where a heat current generates a charge potential, we demonstrate that a spin current in a paramagnet leads to a charge potential, if the conductivity is energy dependent. We use graphene \cite{castro-neto_electronic_2009} as a model system to study this effect, as recently proposed \cite{vera-marun_non-linear_2011}. The physical concept demonstrated here is generally valid, opening new possibilities for spintronics.
}

Previous reports on detection of spin signals using non-magnetic contacts have made use of spin-orbit interaction via the (inverse) spin Hall effect \cite{valenzuela_direct_2006, uchida_observation_2008}. Recently, large non-local signals in graphene have been attributed to an effect with similar phenomenology, given by the difference in Hall resistance between two (spin) channels induced by an applied perpendicular magnetic field \cite{abanin_giant_2011-3}. Both effects produce a charge potential transversal to the direction of the spin current and are valid in the linear regime. In the present work we deal with a different concept based on a non-linear interaction between spin and charge which results in charge potentials longitudinal to the spin current \cite{vera-marun_non-linear_2011}. This effect is solely based on the energy dependence of the conductivity $\sigma(\epsilon)$, not requiring spin-orbit interaction nor external magnetic fields.

\begin{figure}[tbp] %width=85mm
\includegraphics*[angle=0, width=0.5\textwidth]{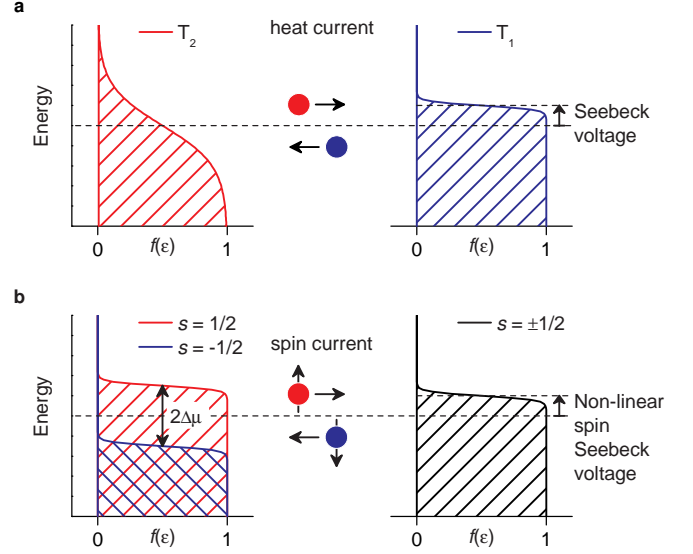}
\caption{\label{fig:concept}
\textbf{Analogy between spin and heat transport illustrated by electronic distributions $f(\epsilon)$.} \textbf{a,} A temperature gradient ($T_2 > T_1$) sets up a heat current, with high-energy electrons moving towards the cold region and low-energy electrons moving towards the hot region. When the conductivity is energy dependent (here $\partial \sigma / \partial \epsilon > 0$) a Seebeck voltage is built up under open circuit conditions, to compensate for the different conductivities of high- and low-energy electrons. \textbf{b,} A gradient in spin accumulation $\Delta \mu$ sets up a spin current, with majority spin electrons moving towards the region with lower $\Delta \mu$ and minority spin electrons moving in the opposite direction. Similar to thermoelectricity, a voltage is built up if the conductivity is energy dependent due to the different conductivities of the two electron spin species.
}
\end{figure}    %\vspace*{5mm}

To explain the concept of detection of spin signals used here it is useful to make an analogy with thermoelectrics. As shown in Fig.~\hyperref[fig:concept]{\ref*{fig:concept}a}, a temperature gradient sets up a heat current. Under open circuit conditions, this results in a built up voltage $V = -S (T_2 - T_1)$, with $S$ the Seebeck coefficient of the conducting system. For the case of diffusive spin transport \cite{valet_theory_1993} (see Fig.~\hyperref[fig:concept]{\ref*{fig:concept}b}) the electrochemical potential of each spin channel can be described as $\mu_\pm=\mu_{\text{avg}}\pm\Delta\mu$, with $\Delta\mu$ the spin accumulation (created by electrical spin injection) in the conductor, which decays with a characteristic spin relaxation length $\lambda$. The gradient in spin accumulation  sets up a spin current which results in a built up voltage $V = -(\beta /e) (\Delta\mu_2 - \Delta\mu_1)$, with $\beta$ the conductivity spin polarization of the system \cite{valet_theory_1993}. While $S$ is a general property of conductors, $\beta$ is in general zero except for ferromagnetic or ferrimagnetic materials. Therefore, pure spin currents are not expected to generate charge voltages in a paramagnet like graphene. The latter is not true if we consider spin transport away from the Fermi level. When a sizable $\Delta\mu$ is present, each spin channel experiences a different conductivity even in a paramagnet, as long as the conductivity is energy dependent. So we consider a spin polarization of the conductivity induced by $\Delta\mu$, which can be approximated as \cite{vera-marun_non-linear_2011} $\beta = - \Delta\mu \sigma^{-1} \partial \sigma / \partial \epsilon$.

\begin{figure}[tbp]
\includegraphics*[angle=0, width=0.5\textwidth]{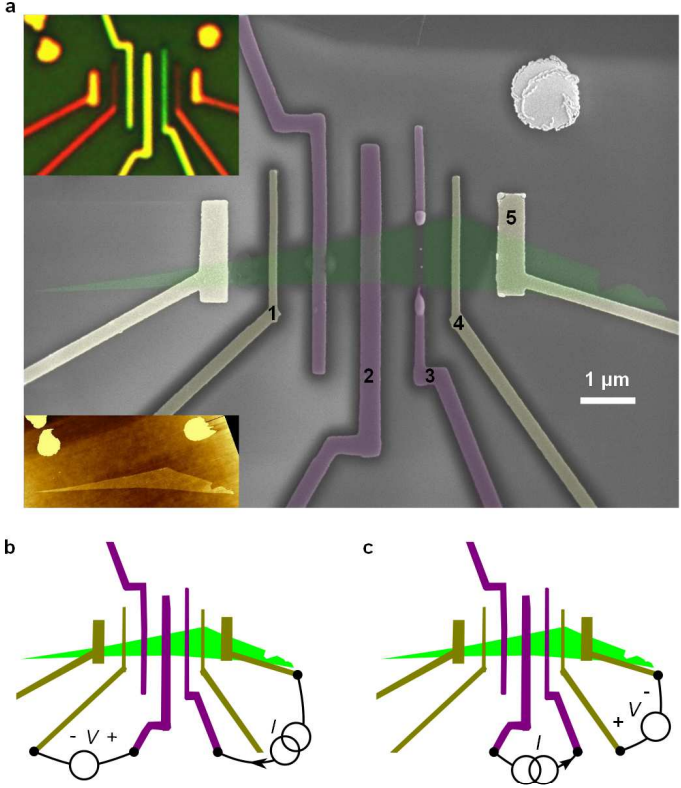}
\caption{\label{fig:sample}
\textbf{Sample geometry and non-local measurement configuration.}
\textbf{a,} Coloured SEM image of the device (after measurement and sample failure). Tunnel contacts have electrodes made of 5/25-nm-thick Ti/Au (Contacts 1, 4 and 5) or 30-nm-thick Co (Contacts 2 and 3). Top inset: optical image before measurement. Bottom inset: AFM image of graphene before contact deposition. \textbf{b,} Configuration for measuring linear spin resistance using a magnetic detector. \textbf{c,} Configuration for measuring non-linear resistance using non-magnetic detectors.
}
\end{figure}

\begin{figure}[tbp]
\includegraphics*[angle=0, width=0.5\textwidth]{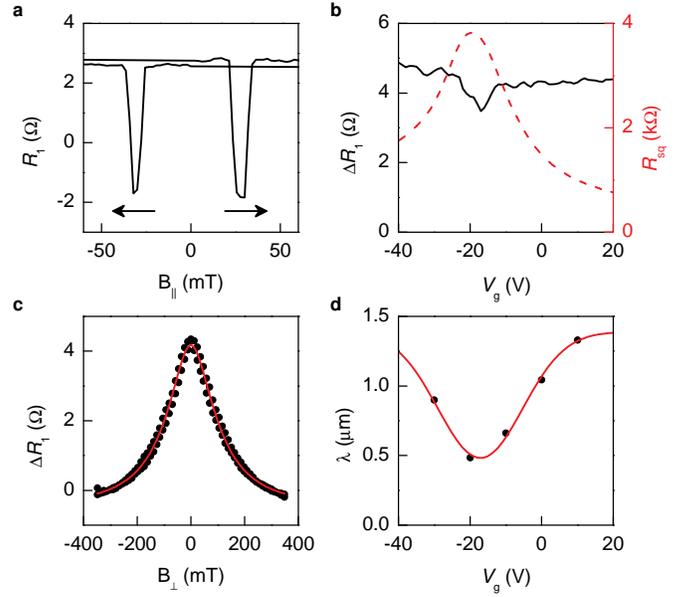}
\caption{\label{fig:Rs}
\textbf{Linear spin detection using a magnetic detector.} \textbf{a,} Spin-valve effect in non-local linear resistance $R_1$ by sweeping an in-plane magnetic field at $V_g = 0$~V. Two well-defined values correspond to parallel ($R_1^{\text{Pa}}$) and anti-parallel ($R_1^{\text{Ap}}$) alignment of Co contacts. \textbf{b,} Spin resistance $\Delta R_1 = R_1^{\text{Pa}} - R_1^{\text{Ap}}$ versus $V_g$. The dashed line is the square resistance $R_{\text{sq}}$ of graphene between Contacts 2 and 3 with $V_D \approx -20$~V. \textbf{c,} Hanle spin precession curve by sweeping a perpendicular magnetic field at $V_g = 10$~V. The solid line is a fit with the one-dimensional Bloch equation. The obtained parameters are $D = 0.025$~m$^2/$s and $\tau = 71$~ps, with contact spin polarization $P = 9$~\%. \textbf{d,} Spin relaxation length $\lambda$ extracted from Hanle curves (as that shown in \textbf{c}) taken at several values of $V_g$. The solid line is a fit with a Gaussian function for parameterization purposes.
}
\end{figure}

To complete the analogy, we define $\alpha = \sigma^{-1} \partial \sigma / \partial \epsilon$ and the Lorentz number $L_0 = (\pi^2/3)(k^2_B/e^2)$. The previous expressions for $V$ are only valid when the coefficients $S$ and $\beta$ are independent of the driving forces $T$ and $\Delta\mu$, respectively. In reality, the Seebeck coefficient is given by the Mott formula \cite{cutler_observation_1969, zuev_thermoelectric_2009} $S = - L_0 e \alpha T$. In the limit $T \approx T_2 - T_1$, the Seebeck voltage depends quadratically on the driving force as $V \propto L_0 e \alpha (T)^2$. Similarly, for spin transport in a paramagnet, the induced spin polarization mentioned above ($\beta = -\alpha \Delta\mu$) also results in a quadratic dependence on the driving force as $V \propto (\alpha / e)(\Delta\mu)^2$. Due to the common factor $\alpha$ the effect described here has similar behavior as the Seebeck coefficient, showing opposite polarity for electron and hole regimes. Furthermore, because $\Delta\mu$ is (to lowest order) linear on the injection current \cite{jedema_electrical_2001, tombros_electronic_2007}, the result is a second order signal $V \propto I^2$.

For our proof of concept we use graphene which, apart from being a two-dimensional platform for relativistic quantum mechanics \cite{castro-neto_electronic_2009}, has proven to be an excellent system for spin transport where large values of $\Delta\mu \approx 1$~meV can be obtained \cite{tombros_electronic_2007}. The sample is shown in Fig.~\hyperref[fig:sample]{\ref*{fig:sample}a}. It consists of a lateral graphene field-effect transistor covered with a thin aluminum oxide barrier that yields high-resistance contacts for efficient spin transport \cite{schmidt_fundamental_2000} and electrostatic gating via the Si/SiO$_2$ substrate, as previously reported \cite{tombros_electronic_2007, popinciuc_electronic_2009}. Besides using magnetic Co contacts for electrical spin injection and detection (see Fig.~\hyperref[fig:sample]{\ref*{fig:sample}b}), we also include Ti/Au contacts. These non-magnetic contacts are used to electrically detect spins in the configuration shown in Fig.~\hyperref[fig:sample]{\ref*{fig:sample}c}. There are two key aspects to such a measurement configuration. First, the use of non-magnetic detectors simplifies the analysis of the non-local signal, because for the case of using magnetic detectors both linear and non-linear signals are expected due to direct detection of $\Delta\mu$ \cite{bakker_interplay_2010, vera-marun_non-linear_2011}. Second, the use of two magnetic contacts as source and drain for charge current allows us to measure the non-local signal for both parallel $V^{\text{Pa}}$ and anti-parallel $V^{\text{Ap}}$ alignment of their magnetizations and thereby to focus on the difference between the two states $\Delta V$. This way we can exclude background signals which do not depend on $\Delta\mu$ and can be present in non-local measurements \cite{bakker_interplay_2010}. We use a lock-in technique to determine the linear $V_1$ and second order $V_2$ components of the resulting root mean square signal. From them we extract the non-local resistances $R_i$ contributing to the total signal $V = R_1 I + R_2 I^2$. All measurements are at room temperature, unless otherwise noted.

\begin{figure}[tbp]
\includegraphics*[angle=0, width=0.5\textwidth]{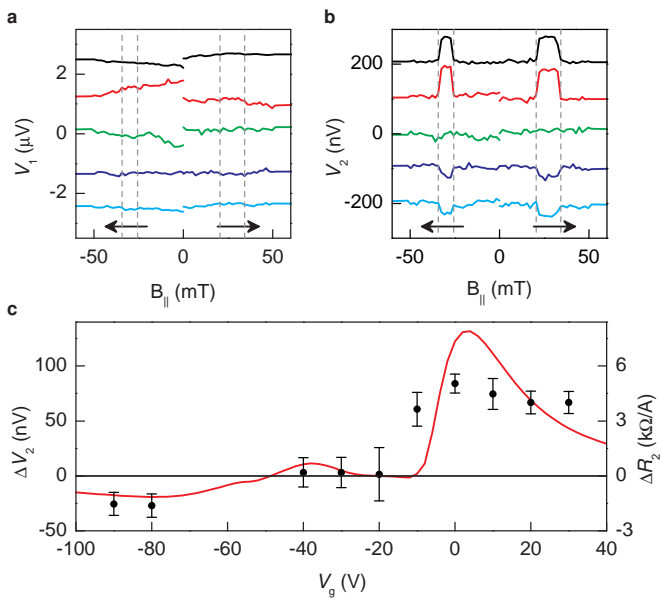}
\caption{\label{fig:R2}
\textbf{Non-linear spin detection using non-magnetic detectors.} \textbf{a,} Linear non-local signal versus in-plane magnetic field, no spin-valve effect is observed. \textbf{b,} Second order signal (same measurement as in \textbf{a}) showing spin-valve effect. Two well-defined values correspond to parallel ($V_2^{\text{Pa}}$) and anti-parallel ($V_2^{\text{Ap}}$) alignment of the Co contacts. Curves in \textbf{a} and \textbf{b} correspond to (from top to bottom) $V_g = 30, 0, -40, -80$ and $-90$~V and are offset vertically for clarity. Each curve is the average of 10 measurements. All data for a root mean square current of $5~\mu$A (configuration as in Fig.~\hyperref[fig:sample]{\ref*{fig:sample}c}). \textbf{c,} Non-linear spin resistance $\Delta R_2 = R_2^{\text{Ap}} - R_2^{\text{Pa}}$ versus $V_g$. For each data point the average value of $V_2$ for (anti)parallel configuration, and their standard deviation, was extracted from curves as those shown in \textbf{b}. The solid line is a result from numerical modelling.
}
\end{figure}

\begin{figure}[tbp]
\includegraphics*[angle=0, width=0.5\textwidth]{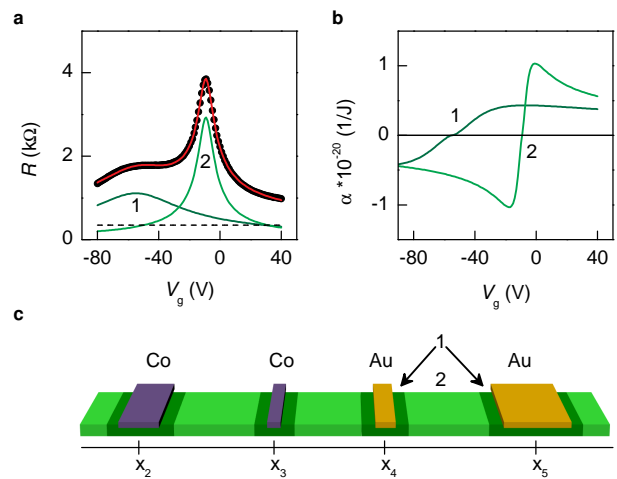}
\caption{\label{fig:alpha}
\textbf{Role of charge density distribution.} \textbf{a,} Resistance of graphene between Contacts 4 and 5 (Au detectors). The red solid line is a fit using a phenomenological description including two Dirac curves (solid lines labeled 1 and 2) and a constant 0.35~k$\Omega$ for the low resistance Contact 5 (dashed line). \textbf{b,} Extracted parameter $\alpha = \sigma^{-1} \partial \sigma / \partial \epsilon$ for each of the two Dirac curves mentioned in \textbf{a}. \textbf{c,} Schematic representation of charge density distribution within graphene. Different properties are considered for graphene under and next to the contacts (1) and for the region away from the contacts (2).
}
\end{figure}

We start by characterizing spin transport in the linear regime. The results in Fig.~\hyperref[fig:Rs]{\ref*{fig:Rs}} show a non-local spin-valve effect, demonstrating spin transport between Contacts 2 and 3, for a center-to-center separation of $L = 1.0~\mu$m and width of graphene $w \approx$~1.1~$\mu$m. The obtained spin resistance $\Delta R_1 \approx 4~\Omega$ is nearly constant versus the gate voltage $V_g$ applied to the substrate. $V_g$ controls the charge carrier density $n_g$ in graphene as $n_g=\gamma(V_g - V_D)$, with $V_D$ the condition for charge neutrality (Dirac point) and $\gamma = 7.2 \times 10^{14}~\text{m}^{-2}\text{V}^{-1}$. On the other hand, the graphene square resistance $R_\text{sq}$ depends on $V_g$, changing by a factor of 5. The observed $\Delta R_1$ vs $R_\text{sq}$ behavior can be understood by the standard relation \cite{tombros_electronic_2007} $\Delta R_1 = (P^2 R_\text{sq} \lambda / w) \exp{(-L/ \lambda)}$, with $P$ the spin polarization of the magnetic contacts, as being due to the charge carrier density dependence of $\lambda$ with a minimum at the Dirac point. The latter is attributed in graphene to the Elliot-Yafet mechanism for spin relaxation \cite{jozsa_linear_2009, avsar_toward_2011, han_spin_2011} (Supplementary A). Furthermore, the previous relation is valid only for contact resistances $R_c \gg R_\text{sq} \lambda /w$, where the contacts do not affect the spin transport \cite{schmidt_fundamental_2000, popinciuc_electronic_2009, han_tunneling_2010}. We take into account both considerations in our modelling below, by parameterizing $\lambda$ (see  Fig.~\hyperref[fig:Rs]{\ref*{fig:Rs}d}) and by including the finite resistance of the contacts used for spin injection and detection.

Next, we demonstrate non-linear detection of spins by using non-magnetic contacts. In Fig.~\hyperref[fig:R2]{\ref*{fig:R2}b} is shown a clear spin-valve effect in the second order component $V_2$ of the non-local signal. The transitions in $V_2$ occur at the switching fields of the magnetic contacts used for current injection. We observe at zero gate voltage that $V_2^\text{Ap} > V_2^\text{Pa}$, consistent with the presence of a larger $\Delta\mu$ for the anti-parallel magnetic configuration \cite{tombros_electronic_2007} and a positive sign of the parameter $\alpha$ for electron transport \cite{vera-marun_non-linear_2011}. Therefore, we expect that the sign of the non-linear spin resistance $\Delta R_2$ should follow that of $\alpha$ and change sign when going from transport in the electron ($V_g > V_D$) to the hole ($V_g < V_D$) regime. The latter is confirmed by observing that $\Delta R_2 < 0$ for gate voltages $V_g \ll V_D$. We remark that the measured spin-valve signal cannot be explained by spurious detection of potentials in the current carrying part of the sample, due to the absence of spin-valve signal in the first order response (Fig.~\hyperref[fig:R2]{\ref*{fig:R2}a}).

The gate voltage dependence of the non-linear spin resistance is presented in Fig.~\hyperref[fig:R2]{\ref*{fig:R2}c}. The $\Delta R_2$ vs $V_g$ curve shows a maximum of $\approx 5$~k$\Omega /$A for electron transport. We did not observe a clear sign change when crossing the Dirac point, whereas for hole transport there was a minimum value of only $\approx -2$~k$\Omega /$A. To understand this electron-hole asymmetry we looked into the charge transport properties of the detector circuit, between Contacts 4 and 5. The Dirac curve in Fig.~\hyperref[fig:alpha]{\ref*{fig:alpha}a} shows that, while there is a reasonable symmetry for $V_g$ close to $V_D = -9$~V, this is not the case for larger $V_g$, as evidenced by the kink visible at $V_g = -55$~V. Such kinks in the Dirac curve have been described as arising due to electron doping from metal contacts having a thin oxide layer that prevents charge density pinning \cite{nouchi_charge-density_2010}. Our contacts are deposited onto a thin oxide barrier. Therefore we interpret the Dirac curve of the detector circuit as being composed of two contributions, the graphene under (and next to) the contacts and that away from the contacts (curves 1 and 2 in Fig.~\hyperref[fig:alpha]{\ref*{fig:alpha}a}, Supplementary B).

Having described both spin and charge transport in the linear regime, we now construct a minimal one-dimensional model that allows quantitative comparison with experiment. As mentioned above, we model $\Delta\mu$ by considering the induced conductivity spin polarization $\beta$, finite resistance contacts, and gate voltage dependency of $\lambda$. We use a fixed $P = 9~\%$ for the magnetic contacts (extracted in the regime $R_c \geq 5 R_{\text{sq}} \lambda /w$) and a width profile for graphene as extracted from atomic force microscopy (AFM). Furthermore, we describe the Dirac curve for each graphene region using the approximate relation \cite{grosse_nanoscale_2011} $\sigma = \nu e (n_g^2+n_i^2)^{1/2}$, with $\nu$ the carrier mobility and $n_i$ a background carrier density due to the presence of electron-hole puddles and thermally generated carriers. We then extract the parameter $\alpha$ for each region \cite{vera-marun_non-linear_2011} (see Fig.~\hyperref[fig:alpha]{\ref*{fig:alpha}b}) likewise to the extraction of the Seebeck coefficient of graphene from the Dirac curve \cite{zuev_thermoelectric_2009, grosse_nanoscale_2011}.

The model, schematically shown in Fig.~\hyperref[fig:alpha]{\ref*{fig:alpha}c}, has the extension of the graphene region 1 beyond the contact edge as the only free parameter. Scanning photocurrent work \cite{mueller_role_2009} has shown that the doping in graphene decays gradually from the contact edge extending up to a distance of $\approx 0.3~\mu$m. For simplicity we consider a constant doping up to a distance of 0.15~$\mu$m. The modelled $\Delta R_2$ vs $V_g$ curve (solid line in Fig.~\hyperref[fig:R2]{\ref*{fig:R2}c}) successfully reproduces both the trend and the magnitude of the data. The agreement yields certainty to our interpretation of the measured $\Delta R_2$ signal as arising due to the non-linear interaction between spin and charge.

The magnitude of $\Delta R_2$ is only slightly limited by the finite resistance of our contacts ($R_c \approx 7~\text{k}\Omega$). Assuming infinite contact resistance, we predict only up to a 2 times increase in $\Delta R_2$. On the other hand, the use of high-quality tunnel contacts \cite{han_tunneling_2010} with $P=30\%$ would yield a 10 times increase. And since $\alpha_\text{max} \propto 1/\sqrt{n_i}$, similar to the Seebeck coefficient \cite{zuev_thermoelectric_2009}, decreasing $n_i$ by two orders of magnitude by using a boron nitride substrate \cite{xue_scanning_2011} would yield a further 10 times increase. Therefore the herewith demonstrated effect is a real candidate for spin detection. It can be regarded as a step in the logical progression from linear interactions between spin and charge towards interactions between spin and heat, as studied in the field of spin caloritronics \cite{bauer_spin_2010}.

%\section*{}
%\section*{References}

\section*{methods}
\subsection*{Sample preparation}
Graphene is obtained from HOPG graphite by mechanical exfoliation and deposited on a highly n-doped Si substrate covered with a thermal oxide layer 300~nm thick. The Si substrate is used as electrostatic gate. Graphene is first covered by 0.8~nm Al followed by natural oxidation to obtain a thin aluminum oxide layer. Electron beam lithography (EBL), deposition by evaporation, and lift-off were performed twice, first for the Ti/Au contacts and then for Co. We observed in previous samples a lower resistance of the Ti/Au contacts, which forbade us to use them for spin detection \cite{schmidt_fundamental_2000}. This problem was possibly related to the required baking of PMMA for the second EBL step, which may cause diffusion of Ti/Au through the oxide barrier. To solve it we increased the thickness of the oxide barrier only for the Ti/Au contacts by deposition and oxidation of extra 0.3~nm Al. This yielded non-invasive Au contacts with similarly high resistances as those of the Co contacts.

\subsection*{Measurements} 
Characterization took place at room temperature and at 77~K (Supplementary A) in a cryostat with a base pressure $\approx 1\times 10^{-7}$~mbar. The sample was first annealed under vacuum at 130$^\circ$C for $\approx$~24~hr for removal of physisorbed water, resulting in low hysteresis in the Dirac curve ($\Delta V_D < 10$~V). Measurements were performed using an a.c.\ current source 1--5~$\mu$A and recording simultaneously the first and second harmonic responses using two lock-in systems. All current and voltage signals reported are root mean squared values. Therefore the resistances were extracted as $R_1 = V_1/I$ and $R_2 = \sqrt{2}V_2/I^2$. Excitation frequency was kept $\leq 3$~Hz to prevent signals due to capacitive coupling, as determined by frequency scans. Contribution from higher harmonics was found to be negligible.

\subsection*{Modelling}
We developed a one-dimensional finite-element code using the program MATLAB to find numerical solutions for the two-channel spin diffusion equations \cite{valet_theory_1993} 
\begin{eqnarray}
\frac{I_\pm}{w} = \sigma_\pm \frac{\partial \mu_\pm}{\partial x} \;, \label{eq:S1}\\
\frac{\partial^2 \Delta\mu}{\partial x^2} = \frac{\Delta\mu}{\lambda^2} \;, \label{eq:S2}
\end{eqnarray}
with the inclusion of an element-specific conductivity spin polarization \cite{vera-marun_non-linear_2011} $\beta$, as described in the main text. Element length was kept $\leq 10$~nm. $I_\pm$ and $\Delta\mu$ were set to be continuous across element boundaries. We consider point contacts located at the center of the fabricated electrodes, which could either inject a spin current $PI$ or detect the electrochemical potential $\mu_{avg} + P \Delta\mu$. Spin relaxation under contacts with finite resistance (4.9, 9.2, 7.3 and $<1$~k$\Omega$ for Contacts 2, 3, 4 and 5, respectively) was implemented as in ref.~\onlinecite{popinciuc_electronic_2009}, with the extra consideration of contact spin polarization $P$. The extracted parameters for the curve Dirac 2 (graphene between contacts) were $\nu = 3900$~cm$^2$/Vs and $n_i = 3.5 \times 10^{15}$~m$^{-2}$, consistent with previous experiments on SiO$_2$ substrates \cite{grosse_nanoscale_2011}, whereas for Dirac 1 we obtained $\nu = 800$~cm$^2$/Vs and $n_i = 2 \times 10^{16}$~m$^{-2}$. To obtain $\Delta R_2$ we calculated the difference in potential $V = -\mu_\text{avg}/e$ between the Au detectors, for both parallel and anti-parallel configurations and d.c.\ currents $I = \pm 5~\mu$A. The result was then fitted with $\Delta V = \Delta R_1 I + \Delta R_2 I^2$. The odd contribution $|\Delta R_1| < 3~\text{m}\Omega$ was found to be negligible.

\section*{Acknowledgements}
We would like to thank J.\ G.\ Holstein, B.\ Wolfs, and M.\ de Roosz for technical assistance.
% and XX for critically reading the manuscript
This work was financed by the Zernike Institute for Advanced Materials and by EU FP7 ICT Grant No.\ 251759 MACALO.

\clearpage

\section*{Supplementary information A}
Here we present spin transport data acquired at liquid-nitrogen temperature (77~K). First, we show linear spin transport at 77~K and compare it with that at room temperature. Then we present partial data on non-linear spin detection using non-magnetic detectors at 77~K.

\begin{figure}[bhp]
\includegraphics*[angle=0, width=0.5\textwidth]{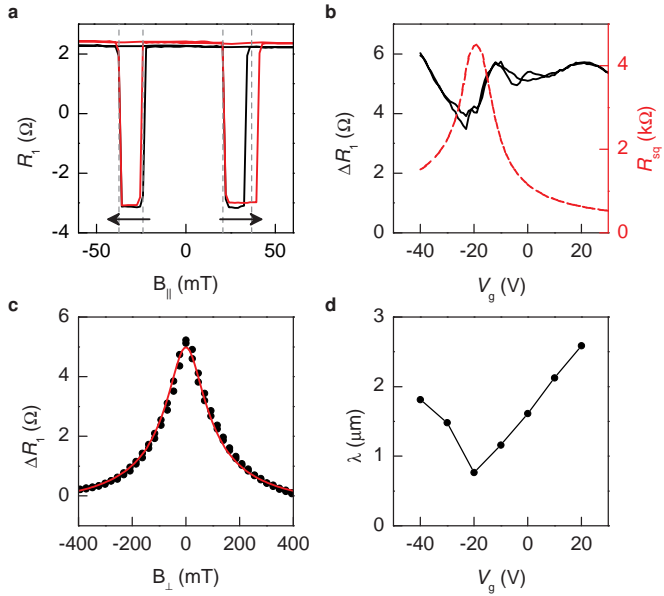}
\caption{\label{fig:suppl-Rs}
\textbf{Linear spin detection using a magnetic detector at 77~K.} \textbf{a,} Two curves showing the spin-valve effect in non-local linear resistance $R_1$ by sweeping an in-plane magnetic field at $V_g = 0$~V. Two well-defined values correspond to parallel ($R_1^{\text{Pa}}$) and anti-parallel ($R_1^{\text{Ap}}$) alignment of Co contacts. \textbf{b,} Spin resistance $\Delta R_1 = R_1^{\text{Pa}} - R_1^{\text{Ap}}$ versus $V_g$. The dashed line is the square resistance $R_{\text{sq}}$ of graphene between Contacts 2 and 3 with $V_D \approx -20$~V. \textbf{c,} Hanle spin precession curve by sweeping a perpendicular magnetic field at $V_g = 10$~V. The solid line is a fit with the one-dimensional Bloch equation. The obtained parameters are $D = 0.053$~m$^2/$s and $\tau = 86$~ps, with contact spin polarization $P = 7$~\%. \textbf{d,} Spin relaxation length $\lambda$ extracted from Hanle curves (as that shown in \textbf{c}) taken at several values of $V_g$.
}
\end{figure}

We start by characterizing spin transport in the linear regime at 77~K. The results in Fig.~\ref{fig:suppl-Rs} show a non-local spin-valve effect, again demonstrating spin transport between Contacts 2 and 3. The spin resistance is $\Delta R_1 \approx 5~\Omega$ and shows a minimum close to the Dirac point. This result for $\Delta R_1$ is similar to that at room temperature (shown in the main text) but $\approx 20~\%$ larger. A larger $\Delta R_1$ at 77~K can be understood from the analysis of Hanle spin precession curves (see Fig.~\hyperref[fig:suppl-Rs]{\ref*{fig:suppl-Rs}c}) from where we extract spin relaxation lengths $\approx 60~\%$ larger than at room temperature. The effect of larger values of $\lambda$ at 77~K is slightly compensated in our sample by a lower contact spin polarization $P = 7~\%$. We remark that the gate voltage dependence of the spin relaxation length $\lambda = \sqrt{\tau D}$ at 77~K (see Fig.~\hyperref[fig:suppl-Rs]{\ref*{fig:suppl-Rs}d}) shows a minimum close to the Dirac point, similar to the data at room temperature. The latter is is indicative of the Elliot-Yafet mechanism of spin relaxation in single-layer graphene \cite{jozsa_linear_2009, avsar_toward_2011, han_spin_2011} being dominant both at room temperature and at 77~K.

\begin{figure}[tbp]
\includegraphics*[angle=0, width=0.5\textwidth]{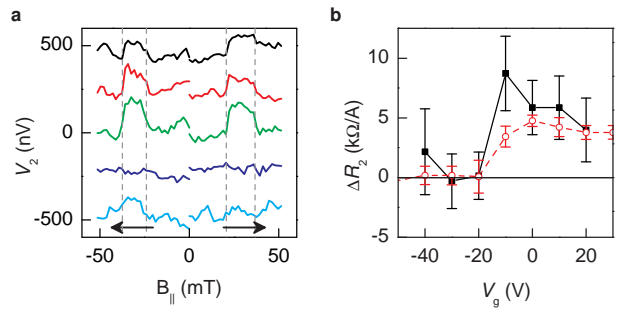}
\caption{\label{fig:suppl-R2}
\textbf{Non-linear spin detection using non-magnetic detectors at 77~K.} \textbf{a,} Second order signal showing spin-valve effect at 77~K. Two well-defined values correspond to parallel ($V_2^{\text{Pa}}$) and anti-parallel ($V_2^{\text{Ap}}$) alignment of the Co contacts. The curves correspond to (from top to bottom) $V_g = 20, 0, -10, -20$ and $-40$~V and are offset vertically for clarity. Each curve is the average of 26 measurements. All data for a root mean square current of $5~\mu$A.
% (configuration as in Fig.~\hyperref[fig:sample]{\ref*{fig:sample}c}).
\textbf{b,} Non-linear spin resistance $\Delta R_2 = R_2^{\text{Ap}} - R_2^{\text{Pa}}$ versus $V_g$ at 77~K (closed black squares). For each data point the average value of $V_2$ for (anti)parallel configuration, and their standard deviation, was extracted from curves as those shown in \textbf{a}. The data for room temperature (open red circles) is also shown for comparison.
}
\end{figure}

Finally, in Fig.~\ref{fig:suppl-R2} we demonstrate non-linear detection of spins by using non-magnetic contacts at 77~K. Although our data at low temperature is limited, it shows a similar behavior of $\Delta R_2$ as that at room temperature. The magnitude of $\Delta R_2$ at both temperatures is similar within the experimental uncertainty, except for an almost 2 times higher value at $V_g = -10$~V close to the Dirac point. The observation of similar results at room temperature and at 77~K are a confirmation of our interpretation of the non-linear spin-valve signal as solely arising from an interaction between spin and charge, which does not directly involve heat as in the case of spin thermoelectric effects \cite{uchida_observation_2008, slachter_thermally_2010-1, le_breton_thermal_2011}.

\section*{Supplementary information B}
Here we discuss on the identification of contributions to the Dirac curve from graphene regions under and around the contacts and those away from the contacts. We also discuss on the nature of the contacts and their possible contributions to the non-linear spin signal.

\begin{figure}[tbp]
\includegraphics*[angle=0, width=0.5\textwidth]{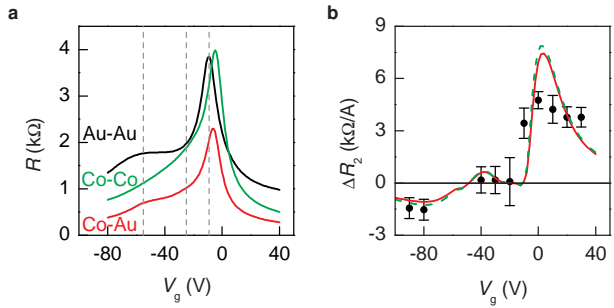}
\caption{\label{fig:suppl-doping}
\textbf{Effect of Co and Au contacts on graphene doping at room temperature.} \textbf{a,} Dirac curve of graphene between Contacts 2 and 3 (Co-Co, green), graphene between Contacts 3 and 4 (Co-Au, red) and graphene between Contacts 4 and 5 (Au-Au, black). Vertical dashed lines indicate the location of the Dirac point $V_D$ for different graphene regions. \textbf{b,} Non-linear spin resistance $\Delta R_2 = R_2^{\text{Ap}} - R_2^{\text{Pa}}$ versus $V_g$. The red solid line is for the model described in the main text, considering graphene regions under both Au and Co contacts to behave the same with $V_D = -55$~V. The dashed green line is for considering graphene under the Co contacts to have a Dirac point $V_D = -20$~V. The data for room temperature (black circles) is also shown for comparison.
}
\end{figure}

In the main text we showed how the Dirac curve for graphene between the two Au detectors 
%(see Fig.~\hyperref[fig:alpha]{\ref*{fig:alpha}a}) 
is composed of two distinct contributions. The main contribution corresponds to regions of the graphene channel located away from the contacts, with a Dirac point $V_D = -9$~V. A minor contribution, visible as a kink in the hole regime \cite{nouchi_charge-density_2010}, corresponds to regions of graphene located under (and next to) the Au contacts with $V_D = -55$~V (due to contact doping). We also observed similar kinks for the Dirac curves for graphene between the adjacent Co injector and Au detector, and for graphene between the two Co contacts used for spin injection (see Fig.~\hyperref[fig:suppl-doping]{\ref*{fig:suppl-doping}a}). The kinks in the Dirac curves indicate that the Co contacts also dope the graphene channel but with a Dirac point close to $V_D = -20$~V, different than for graphene around the Au contacts ($V_D = -55$~V).

The resulting $\Delta R_2$ for the model presented in the main text 
%(see Fig.~\hyperref[fig:R2]{\ref*{fig:R2}c}) 
corresponds to the simple case of assumption that all contacts have the same effect on graphene, with $V_D = -55$~V. In Fig.~\hyperref[fig:suppl-doping]{\ref*{fig:suppl-doping}b} we also show the result of incorporating in the model a different contribution from the Co contacts, with $V_D = -20$~V. Notice this consideration does not have a significant effect on the modelled $\Delta R_2$. There are two reasons for this observation. First, the graphene regions modified by the presence of the Co contacts are not within the detector circuit. Therefore, charge potentials generated due to their $\alpha$ parameter have no influence on the signal detected between the Au contacts. Second, though the graphene regions under the Co contacts do have an influence on the $\Delta \mu$ profile via their resistivity (Dirac curve), this influence is small because these regions are narrow compared to the full extent of graphene over which $\Delta \mu$ decays. So the consideration of doping effects under the Co contacts is not critical for understanding the non-linear spin signal measured via the Au contacts.

A fundamental question is whether the contacts themselves contribute to the measured non-linear spin signal. This signal, generated via the non-linear interaction between spin and charge, relies on achieving a large enough $\Delta\mu$ and having a sizable $\alpha$ parameter. Owing to the large conductivity of metals, the achieved spin accumulation within the Au and Co metals ($\approx 1~\mu$eV) \cite{jedema_electrical_2001} is much lower than in graphene. So we do not expect a sizable signal coming from the bulk of the metallic contacts.

\begin{figure}[tbp]
\includegraphics*[angle=0, width=0.5\textwidth]{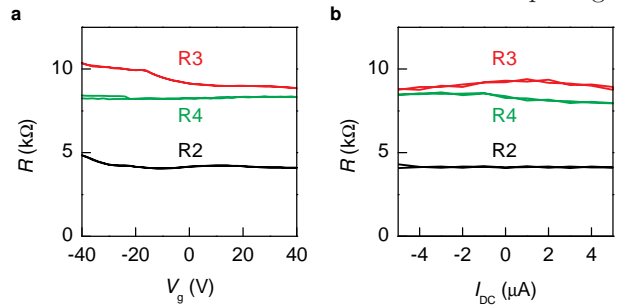}
\caption{\label{fig:suppl-contacts}
\textbf{Contact characterization at 77~K.} \textbf{a,} Gate voltage dependence of resistances of Contacts 2, 3 and 4. Data for a root mean square current of 2~$\mu$A. \textbf{b,} Differential resistances of Contacts 2, 3 and 4 versus d.c.\ current bias, for a root mean square modulation of 0.1~$\mu$A.
}
\end{figure}

The discussion above leaves us with the final possibility that the graphene-metal interface could produce a sizable  signal. Spin thermoelectric effects have been observed in high-quality tunnel contacts \cite{le_breton_thermal_2011}, as expected from the strong energy dependence of electron transmission through a tunnel barrier. To address this issue we have characterized the charge density and bias dependence of contact resistances in our device. From the results in Fig.~\ref{fig:suppl-contacts} we observe that the contacts only change up to $20~\%$ with gate voltage, and have linear $I-V$ characteristics (constant $dV/dI$ within $10~\%$ for the explored biasing currents). These contact characteristics have been previously observed on similar samples and were ascribed to transport dominated by relatively transparent regions in the oxide barrier \cite{popinciuc_electronic_2009}. In this case we do not expect that the interface would exhibit a sizable $\alpha$ parameter and its contribution to the non-linear spin signal would be negligible. We conclude the latter is applicable to our device, as we did not require to include this effect in our model in order to achieve a satisfactory description of the experimental data.

\end{document}